\documentclass[aps,showpacs,amssymb,superscriptaddress,nofootinbib,10pt]{revtex4}
\usepackage{graphicx}
\usepackage{subfigure}
\usepackage{ulem}
\usepackage{amsmath,amssymb,ascmac,color}
\usepackage{type1cm}
\usepackage{bm}

\newcommand\beq{\begin{equation}}
\newcommand\eeq{\end{equation}}
\newcommand\beqa{\begin{eqnarray}}
\newcommand\eeqa{\end{eqnarray}}

\newcommand{\ds}[1]{#1 \hspace{-0.5em}/}  

\newcommand\btau{\mbox{\boldmath$\tau$}}

\begin{document}
\title{Axial anomaly and nesting  in the inhomogeneous chiral phase}

\author{R. Yoshiike and T. Tatsumi}
\affiliation{Department of Physics, Kyoto University, Kyoto 606-8502, Japan}




\begin{abstract}
Axial anomaly and nesting is elucidated in the context of the inhomogeneous chiral phase. 
Using the Gross-Neveu models in 1+1 dimensions, we shall discuss axial anomaly and nesting from two different points of view:
one is homogeneous chiral transition and the other is the Ferrel-Fulde-Larkin-Ovchinnikov (FFLO) state in superconductivity, which  
are closely related to each other by way of duality. 
It is shown that axial anomaly leads to a particular kind of the FFLO state within the two dimensional Nambu-Jona Lasinio model, 
where axial anomaly is manifested in a different mode.  Nesting is a driving mechanism for both phenomena, but its realization has different features. 
We reconsider the effect of nesting in the context of duality.      

\end{abstract}


\maketitle

\section{Introduction}

Nowadays understanding of the QCD phase diagram is one of the main subjects in nuclear physics. 
There have been devoted many efforts to reveal characteristic properties of QCD matter at finite density or temperature theoretically or experimentally \cite{fukushima}.
Among them the behavior of chiral symmetry has attracted much interest at high temperature or density,
since it plays important roles in the vacuum; 
it gives rise to mass generation for hadrons by way of spontaneous symmetry breaking (SSB) or governs low energy hadron dynamics in a model-independent way. 
Recent lattice QCD simulations have suggested that $q{\bar q}$ scalar condensate is reduced
and chiral symmetry is suggested to be restored at high temperature, 
and many model calculations have shown that it is also restored at high density, by using effective models such as the Nambu-Jona Lasinio (NJL) model or Schwinger-Dyson approach \cite{fis}. 
This phenomenon is called chiral transition.

In most studies the uniform scalar condensate has been assumed as an order parameter.
Recently, a possibility of inhomogeneous chiral transition has been suggested, where the order parameter has a spatial modulation \cite{chi,dcdw,nic,buballa}.
The inhomogeneous chiral phase (iCP) has a spatially modulating order parameter $M(z)$ given by the quark condensate, e.g., $M(z)=\langle {\bar q}q\rangle-i\langle{\bar q}i\gamma_5\tau_3 q\rangle\equiv \Delta(z){\rm exp}(i\theta(z))$, for one-dimensional modulation within $SU(2)_L\times SU(2)_R$ symmetry.
Using the Nambu-Jona Lasinio (NJL) model within the mean-field approximation in the infinite $N_c$ limit,
it has been shown that the tricritical point for the chiral transition is replaced by the Lifshitz point, from which the three phases, the SSB phase, chiral-restored phase and iCP diverge.
Various forms of the spatial modulation are possible, and dual chiral density wave (DCDW) and real kink crystal (RKC) are typical examples;
the former is specified by the complex order parameter, $\Delta(z)=\lambda$ and $\theta(z)=qz$ with the wave vector $q$, while the latter by the real order parameter, $\Delta(z)={\tilde\lambda}\sqrt{\nu}{\rm sn}({\tilde\lambda}z;\nu)$ with modulus $\nu$ and $\theta(z)=0$.

Such transition may give an important impact on the theoretical studies of the QCD phase diagram or some observations in high-energy heavy-ion collisions or compact stars; the Lifshitz point in the QCD phase diagram, spontaneous magnetization of iCP \cite{yoshiike1}, or solidification of quark matter is an example.
 
It is well-known that nesting plays an important role for the appearance of iCP.
Moreover, axial anomaly also plays an important role in some situations, e.g. in the presence of the magnetic field \cite{tatsumi,son}.
We elucidate how these concepts play for inhomogeneous chiral transition.
We consider here iCP in 1+1 dimensions by using the effective models to clearly see their interplay.
We know that iCP with one dimensional modulation in 1+3 dimensions can be studied by using the results obtained in 1+1 dimensions,
and some characteristic features can be discussed by referring to the 1+1 dimensional models \cite{basar}.
Actually DCDW or RKC can be obtained by boosting the general solutions known in the NJL$_2$ model.
Manifestation of nesting or axial anomaly should be also a common feature of iCP in any dimension,
since these concepts are based on geometry of the Fermi surface and chiral symmetry itself.

We shall figure out the characteristic roles of axial anomaly and nesting and their interplay in the context of iCP. 
We use the duality transformation for this purpose.     
Thies has pointed out that there is a duality relation between chiral transition and a kind of superconductivity \cite{thies03s}. 
Actually we can see that chiral condensate is mapped into the Cooper pair condensate.
Then chemical potential can be regarded as an effective magnetic field.
It is well known in condensed matter physics that the BCS state changes to the another state, called as the Fulde-Ferrell-Larkin-Ovchinnikov (FFLO) state, beyond the lower critical field, where the the Cooper pair condensate is spatially modulating \cite{fflo,sar,fflo2,shimahara}.
Thus inhomogeneous chiral transition at finite density is mapped into the problem of the FFLO state in a kind of superconductivity in the vacuum under the external magnetic field. 

First we discuss how axial anomaly is mapped by the duality transformation.
We shall see a new kind of anomaly manifested in a superconducting model, which exhibits an interesting phase diagram and very different from the FFLO state in the condensed matter physics: the Fulde-Ferrell (FF) state may be realized even in a tiny magnetic field.
We figure out how such difference appears by comparing the model with an anomaly free model, two flavor NJL$_2$ model.
Next we figure out the important role of nesting.
Nesting is one of the key mechanism for spatially inhomogeneous phases such as charge density wave, spin density wave in quasi-one dimensional systems in condensed matter physics \cite{sdw,cdw,grud} , and pion condensation in nuclear matter \cite{mig};
the energy gap is opened at the Fermi surface
and the spatial modulation of the order parameter is characterized by the order of twice the Fermi momentum.
It has been sometimes discussed that nesting is responsible for the appearance of DCDW or its one dimensional analog, chiral spiral,
since the wave number $q$ always takes $2\mu$.
On the other hand, the wave number of RKC begins with $q=0$.
Nesting of the Fermi surface is one of the essential ideas in condensed matter physics
and it leads to charge density wave or spin density wave in quasi-one dimensional systems.
Since iCP may be regarded as a generation of a kind of density wave,
we would like to look into nesting in the context of iCP in detail. 
We shall see how manifestation of nesting is changed after the duality transformation,
and nesting in RKC may be clearly seen in the Larkin-Ovchinnikov (LO) state.

In sec. II we briefly review the role of axial anomaly in the context of chiral spiral. 
Another manifestation or mapping of anomaly is discussed in sec III after the duality transformation. Introducing two kinds of nesting, we elucidate the effect of nesting in the iCP in sec IV. Sec. V is devoted to summary and concluding remarks. 

\section{Axial anomaly in chiral spiral}

Here we briefly review how axial anomaly plays a role in iCP by using the Gross-Neveu (GN) models.  
There are various versions of the Gross-Neveu models in 1+1 dimensions with either discrete or continuous chiral symmetry. Among them the chiral GN model or the two-dimensional Nambu-Jona Lasinio model (NJL$_2$) is the most popular version with continuous chiral symmetry,
\beq
{\cal L}_{\rm NJL_2}={\bar \psi}i\ds\partial\psi+\frac{G}{2}\left[\left({\bar\psi}\psi\right)^2+\left({\bar\psi}i\gamma_5\psi\right)^2\right],
\label{njl2}
\eeq
which is invariant under $U(1)_L\times U(1)_R$. This Lagrangian is one-flavor case, but easily extended to the $N$ flavor case endowed with $SU(N)_L\times SU(N)_R$ symmetry. For two-flavor case, 
it renders
\beq
{\cal L}_{\rm 2fNJL_2}={\bar \psi}i\ds\partial\psi+\frac{G}{2}\left[\left({\bar\psi}\psi\right)^2+\left({\bar\psi}i\gamma_5\btau\psi\right)^2\right], 
\label{2fnjl}
\eeq
which we call the 2fNJL$_2$ model.

Both models exhibit spontaneous chiral symmetry breaking. There have been many studies about iCP by using the NJL$_2$ model \cite{basar,ohwa},
and little has used ${\cal L}_{\rm 2fNJL_2}$ to study flavor asymmetric matter \cite{ebe,ebert1,thies16}.
Chiral spiral is defined by $M(x)=\langle {\bar \psi}\psi\rangle-i\langle{\bar \psi}i\gamma_5 \psi\rangle\equiv \Delta(x){\rm exp}(iqx)$ in 1+1 dimensions, and the most favorable phase on the $T-\mu$ plane within the NJL$_2$ model.
The wave vector $q$ then satisfies the relation, $q=2\mu$, which looks to be the same as the nesting vector in spin density wave or charge density wave in quasi-one dimensional systems in condensed matter physics \cite{cdw,sdw}.
Accordingly it has been sometimes discussed that chiral spiral is caused by the nesting effect of the Fermi surface.
When we consider DCDW in 1+3 dimensions, it appears with the wave vector of $O(\mu)$ \cite{dcdw}.
This phenomenon may be understood as a reminiscence of the nesting effect. 

It is to be noted that the effect of anomaly should plays an important role in 1+1 dimensions, without any gauge field. 
Introducing a fictious gauge field $B_\mu$, $B_\mu=(\mu,0)$, we can consider QCD in the background of $B_\mu$. It has been shown that axial-vector current $j^\mu_5={\bar\psi}\gamma^\mu\gamma_5\psi$ is not conserved by anomaly,
\beq
\partial_\mu j^\mu_5=\frac{1}{2\pi}\epsilon^{\mu\nu}B_{\mu\nu},
\label{ano}
\eeq  
for one flavor case, where $B_{\mu\nu}=\partial_\mu B_\nu-\partial_\nu B_\mu$ is the field tensor. 
This anomaly is an analog of axial anomaly in the presence of the electromagnetic field \cite{son}, and it is easily extended for the 1+3 dimensional case, e.g. 
in the presence of the magnetic field.

Adding a proper term for chemical potential $\mu$,  we have an effective Lagrangian,
\beq
{\cal L}_{\rm MF}={\bar \psi}i\ds\partial\psi-m{\bar\psi}{\rm exp}(-i\gamma_5 qx)\psi+\mu{\bar\psi}\gamma_0\psi, \label{lag1f}
\eeq
within the mean-field approximation, where $m$ means the dynamical mass, $me^{iqx}=-G(\langle{\bar\psi}\psi\rangle - i\langle{\bar\psi}i\gamma^5\psi\rangle)$.
Using the Weinberg transformation such that $\psi_W={\rm exp} (-i\gamma_5 qx/2)\psi$, we have 
\beq
{\tilde{\cal L}_{\rm MF}}={\bar \psi_W}i\ds\partial\psi_W-{\bar\psi_W}\left[m+\gamma_0 q/2 \right]\psi_W+\mu{\bar\psi_W}\gamma_0\psi_W
\label{lagifmf}
\eeq
Usually quark number becomes a finite value once $\mu$ is greater than the dynamical mass $m$ at $T=0$. 
However, chiral spiral phase develops from $\mu=0$ due to axial anomaly: 
the single-particle energy is given by $\varepsilon_k=\pm (k^2+m^2)^{1/2}+q/2$ and the energy spectrum is shifted by $q/2$ from the free one. 
Anomalous quark number density is then generated by the spectral asymmetry and is closely related to axial anomaly \cite{niemi}.
The quark number is defined by using the Atiyah-Padori-Singer $\eta$ invariant $\eta_H$,
\beqa
n &=& \frac{1}{2} \int \frac{dx}{L} \langle [\psi^\dagger,\psi] \rangle \nonumber \\
&=& -\frac{1}{2}\eta_H + \sum_k \left[ \theta(\varepsilon_k)n_F(\varepsilon_k-\mu) - \theta(-\varepsilon_k)n_F(-\varepsilon_k+\mu) \right]
\label{num}
\eeqa
with
\beq
\eta_H= \lim_{s\rightarrow +0} \sum_k{\rm sign}(\varepsilon_k)|\varepsilon_k|^{-s}, \label{eta}
\eeq
where $n_F(\varepsilon)=(1+e^\varepsilon)^{-1}$ is the Fermi-Dirac distribution function, $\eta_H$ is proportional to $q$, and the particle number is not necessarily zero for any chemical potential \cite{tatsumi}.
That is why chiral spiral develops from $\mu=0$.


It should be interesting to see that there is no anomaly for ${\cal L}_{\rm 2fNJL_2}$. The anomaly relation (\ref{ano}) can be easily extended to the two flavor case: for the axial-vector current $j_5^\mu={\bar\psi}(\tau_3/2)\gamma^\mu\gamma_5\psi$,
\beq
\partial_\mu j_5^\mu=\frac{1}{4\pi}{\rm tr}(\tau_3)\epsilon^{\mu\nu}B_{\mu\nu}=0,
\eeq
where the fictious gauge field $B^\mu$ should read $B^\mu=(\mu,0)$ with $\mu_u=\mu_d=\mu$ in flavor symmetric matter. The chiral spiral is defined as 
$me^{iqx}=-G(\langle{\bar\psi}\psi\rangle - i\langle{\bar\psi}i\gamma^5\tau_3\psi\rangle)$  in this case and the effective Lagrangian renders
\beq
{\cal L}_{\rm 2fMF}={\bar \psi}i\ds\partial\psi-m{\bar\psi}{\rm exp}(-i\gamma_5\tau_3 qx)\psi+\mu{\bar\psi}\gamma_0\psi
\eeq
under the mean-field approximation. Accordingly the Weinberg transformation is modified to $\psi_W={\rm exp} (-i\gamma_5\tau_3 q/2)\psi$ and we find 
\beq
{\tilde{\cal L}_{\rm 2fMF}}={\bar \psi_W}\left[i\ds\partial-m-\gamma_0\tau_3 q/2 \right]\psi_W+\mu{\bar\psi_W}\gamma_0\psi_W.
\eeq
The single-particle energy is now flavor dependent: $\varepsilon_u=\pm (p^2+m^2)^{1/2}+q/2$ and $\varepsilon_d=\pm (p^2+m^2)^{1/2}-q/2$. Thus the energy spectrum of $u$ quarks is shifted upward by $q/2$, while the one of $d$ quarks is shifted downward by $q/2$ from the free case. Consequently, the spectral asymmetry of $u$ and $d$ quarks cancel each other and leave no anomalous quark number.
Since the wave number may be regarded as an ``isospin chemical potential", $\mu_3=-q/2$, in this case, we study the phase diagram for given $\mu$ by changing $\mu_3$. 




Thus chiral spiral appears above $\mu_c\simeq0.68$ in the 2fNJL$_2$ model \cite{ebert1}, in contrast with the NJL$_2$ model.
It is interesting to see some similar feature to RKC, which also appears above the critical chemical potential $\mu_c=2/\pi$ in the NJL$_2$ model \cite{basar}. 
Since there is no axial anomaly for both RKC and chiral spiral within the 2fNJL$_2$ model,
the phase boundaries between iCP and the chiral-restored phase are identical.
Actually, it should be determined by the correlation function in the chiral-restored phase regardless of the detail of the inhomogeneous condensate \cite{yoshiike3}.

\section{Mapping of anomaly through the duality transformation}
\subsection{NJL$_2$ case}

We now consider another manifestation of axial anomaly in the context of iCP.
Thies have shown that there is a duality between chiral transition and a kind of superconducting models \cite{thies03s}, using the NJL$_2$ model. Duality transformation is, $\psi \rightarrow \chi = \frac{1}{2}(1-\gamma^5)\psi + \frac{1}{2}(1+\gamma^5)\psi^*$, where $^t\psi = (\psi_R, \psi_L)$.
This is a canonical transformation and  the Eq.\,(\ref{njl2}) can be written as 
\beq
{\cal L}_{1f} = {\bar \chi}i\ds\partial \chi + \frac{G}{2} \left(\bar{\chi}^c \chi \right) \left( \bar{\chi}^c \chi \right)^\dagger \label{cooper}
\eeq
by introducing new fields, $\chi_L=\psi_L, \chi_R=\psi_R^*$, in terms of left-handed (L-) and right-handed (R-) Weyl fields, $^t\chi=(\chi_R,\chi_L)$.
$\chi^c$ denotes the charge conjugation field, $\chi^c \equiv \gamma^5\chi^*$.
The Lagrangian is called the Cooper pair model, which is a toy model of the color superconductivity \cite{chodos}.
For the chemical potential term, it is changed to,
\beq
\delta L=\mu\left(\chi_L^*\chi_L - \chi_R^*\chi_R\right),
\eeq
which resembles the interaction term between ``magnetic field" $\mu$ and ``spin-up" (R-) and ``spin-down" (L-) quarks, or $\mu$ may be regarded as ``chiral chemical potential $\mu_5$".
In the following we use the notation $h$ instead of $-\mu$.
Considering the pairing between left-handed and right-handed quarks,
the Hamiltonian within the mean field approximation renders,
\beqa
 H_{1f} &=& \frac{1}{2} \int dx \left[ \chi^\dagger(\gamma^5\hat{p}-\gamma^5h)\chi + \chi^{c\dagger}(\gamma^5\hat{p}+\gamma^5h)\chi^c + \Delta^*\bar{\chi}^c\chi + \Delta\bar{\chi}  \chi^c + \frac{|\Delta|^2}{G} \right]  \nonumber \\
&=& \int dx \left[\Psi^\dagger \left( \hat{p} \sigma_3 + h + \sigma_1{\rm Re}\Delta + \sigma_2{\rm Im}\Delta \right)\Psi +\frac{|\Delta|^2}{2G}\right], \label{ham1f}
\eeqa
with the choice of the Dirac matrices as $\gamma_0=\sigma_1, \gamma^1=-i\sigma_2$ and $\gamma_5=\sigma_3$, where $\Psi^\dagger = (\chi_R, \chi^*_L)$.
The gap equation takes the form,
\beqa
 \Delta = - \frac{G}{2} \langle \bar{\chi}^{c} \chi \rangle.
\label{gap}
\eeqa
Under the duality transformation the chiral condensate made of quark-anti-quark is transformed to the Cooper pair condensate in the context of superconductivity.
Thus chiral transition on the  $T-h$ plane is mapped into superconducting transition under the magnetic field in the vacuum.
If the Cooper pair condensate is spatially modulating,
such phase can be described as the FFLO phase.

We can see how axial anomaly inherent in the Lagrangian (\ref{njl2}) is mapped into the Lagrangian (\ref{cooper}), following ref.\,\cite{son}. 
Since the phase of the gap function defined in Eq.\,(\ref{gap}) represents the phonon degree of freedom $\varphi$, 
it transforms as $\varphi\rightarrow \varphi+2\alpha$ under the $U(1)$ transformation, $\chi\rightarrow {\rm exp}(i\alpha)\chi$. 
In the presence of a fictious axial-vector gauge field, 
$C_\mu=(h,0)$, we have an anomaly relation for the vector current $j^\mu={\bar\chi}\gamma^\mu\chi$ by way of the vacuum polarization,
\beq
\partial_\mu j^{\mu} =\frac{1}{2\pi}\epsilon_{\mu\nu}C^{\mu\nu},
\eeq 
with the field strength, $C_{\mu\nu}=\partial_\mu C_\nu-\partial_\nu C_\mu$. This is an analog of Eq.~(\ref{ano}). Accordingly the effective action changes,
\beq
\delta S=-\int dx\partial_\mu\alpha j^\mu.
\label{ano}
\eeq
Thus the effective Lagrangian must include the relevant term, ${\cal L}_{\rm ano}=\frac{1}{2\pi} \frac{d\varphi}{dx} h$, by way of anomaly matching, and the coefficient of $h$ may be regarded as magnetization.

As the case with anomaly in the superconducting states, we consider the FF state under the magnetic field in the vacuum by using the Eq.\,(\ref{ham1f}), where $\Delta = me^{-iqx}$ is assumed.
The Hamiltonian can be rewritten by the Nambu-Gorkov formalism,
\beqa
 H_{1f} &=& \frac{1}{2} \int dx \left[ (\chi^\dagger,\chi^{c\dagger})
 \begin{pmatrix}
 -i\gamma^5\partial_x - \gamma^5h & \gamma^0 me^{-iqx} \\
 \gamma^0 me^{iqx} & -i\gamma^5\partial_x + \gamma^5h
 \end{pmatrix}
 \begin{pmatrix}
 \chi \\
 \chi^c
 \end{pmatrix} + \frac{m^2}{G}  \right]\nonumber \\
 &=& \frac{1}{2} \int dx \left[ (\chi'^\dagger,\chi'^{c\dagger})
 \begin{pmatrix}
 -i\gamma^5\partial_x - \gamma^5(h - q/2) & \gamma^0 m \\
 \gamma^0 m & -i\gamma^5\partial_x + \gamma^5 (h - q/2)
 \end{pmatrix}
 \begin{pmatrix}
 \chi' \\
 \chi'^c
 \end{pmatrix} + \frac{m^2}{G} \right], \label{ham-NG}
\eeqa
where $\chi'\equiv e^{-iqx/2}\chi$.
The fermion fields are expanded as a series of the eigenstates,
\beqa
 \chi(x) &=& \int \frac{dp}{2\pi}e^{i(p+q/2)x} \frac{1}{\sqrt{2\epsilon_p}}
 \begin{pmatrix}
  \alpha_p\sqrt{\epsilon_p + p}  + \beta^{c\dagger}_p\sqrt{\epsilon_p - p} \\
  \beta_p\sqrt{\epsilon_p - p}  - \alpha^{c\dagger}_p\sqrt{\epsilon_p + p}
 \end{pmatrix}, \label{chi} \\
 \chi^c(x) &=& \int \frac{dp}{2\pi}e^{i(p-q/2)x} \frac{1}{\sqrt{2\epsilon_p}}
 \begin{pmatrix}
  \beta_p\sqrt{\epsilon_p + p}  + \alpha^{c\dagger}_p\sqrt{\epsilon_p - p} \\
  \alpha_p\sqrt{\epsilon_p - p}  - \beta^{c\dagger}_p\sqrt{\epsilon_p + p} 
 \end{pmatrix}, \label{chic}
\eeqa
where $\epsilon_p = \sqrt{p^2 + m^2}$, and, $\alpha_p, \beta_p, \alpha^c_p, \beta^c_p$, are annihilation operators of the quasiparticles after the Bogoliubov transformation.
However, the four annihilation operators are not independent;
there is the relation, $\alpha_p(\beta_p) = \alpha_{-p}^c(\beta_{-p}^c)$, because they must satisfy the relation, $\chi^c = \gamma^5\chi^*$.
Accordingly there appear four branches in the energy spectrum, 
\beqa
 E_\alpha &=& \epsilon_p - h + q/2, \nonumber \\
 E_\beta &=& \epsilon_p + h - q/2, \nonumber \\
 E_\alpha^c &=& -\epsilon_p + h - q/2, \nonumber \\
 E_\beta^c &=& -\epsilon_p - h + q/2. \label{spect}
\eeqa
The ground state is then defined by filling the negative energy states:
\beqa
 \alpha_p |\sigma\rangle &=& 0 ~(E_\alpha>0), \nonumber\\
 \beta_p |\sigma\rangle &=& 0 ~(E_\beta>0), \nonumber \\
 \alpha_p^\dagger |\sigma\rangle &=& 0 ~(E_\alpha<0), \nonumber \\
 \beta_p^\dagger |\sigma\rangle &=& 0 ~(E_\beta<0). \label{sigma}
\eeqa


Since the energies of the quasiparticles (\ref{spect}) exhibits spectral asymmetry, one may expect anomalous particle number as in Eq.\,(\ref{eta}). 
However, we can see that it never induces anomalous particle number, different from the Lagrangian (\ref{lag1f}).
Note that the number of particles is not identical with that of quasiparticles due to the Bogoliubov transformation \cite{tin}.
The particle number density can be evaluated in the same manner as in Eq.\,(\ref{num}) and we find,
\beqa
 n &\equiv& \frac{1}{2} \int \frac{dx}{L} \langle \sigma | [\chi^\dagger, \chi] | \sigma \rangle \nonumber \\
 &=& \lim_{\Lambda \to \infty} \langle \sigma | \int_{-\Lambda-q/2}^{\Lambda-q/2} \frac{dp}{2\pi} \left[ \left( \alpha^\dagger_p \alpha_p - \alpha^\dagger_{-p} \alpha_{-p} \right) \frac{\epsilon_p + p}{2\epsilon_p} + \left( \beta^\dagger_p \beta_p - \beta^\dagger_{-p} \beta_{-p} \right) \frac{\epsilon_p - p}{2\epsilon_p} \right] | \sigma \rangle \nonumber \\
 &=& 0.
\eeqa
In the above calculation, we have used the relation,
\beqa
 \langle \sigma |\alpha^\dagger_p \alpha_p | \sigma \rangle &=& \langle \sigma |\alpha^\dagger_{-p} \alpha_{-p} | \sigma \rangle = n_F(E_\alpha), \\
 \langle \sigma |\beta^\dagger_p \beta_p | \sigma \rangle &=& \langle \sigma |\beta^\dagger_{-p} \beta_{-p} | \sigma \rangle = n_F(E_\beta).
\eeqa
Furthermore, in the limit:\,$m\rightarrow 0$, any physical quantity calculated from the fermion fields (\ref{chi}-\ref{chic}) should coincide with the one in the no interacting case  even if $q$ is still finite.
This is because a physical quantity does not depend on $q$ in the limit, $m\rightarrow 0$,
where the wave number $q$ becomes a redundant variable due to the amplitude $m$ vanishing.
To satisfy the requirement, we need to employ the asymmetric cutoff in the momentum integral, $[-\Lambda - q/2,\Lambda - q/2]$, in the $\chi$ sector (see Appendix A for details).

As is inferred from Eq.~(\ref{ano}), we shall see the appearance of the anomalous magnetization instead.
By using the quasiparticle operators, the magnetization can be evaluated as,
\beqa
 M &\equiv& \frac{1}{2} \int \frac{dx}{L} \langle \sigma | [\chi^\dagger, \gamma^5 \chi] | \sigma \rangle \nonumber \\
 &=& \lim_{\Lambda \to \infty} \langle \sigma | \int_{-\Lambda-q/2}^{\Lambda-q/2} \frac{dp}{2\pi} \left[ \left( \alpha^\dagger_p \alpha_p + \alpha^\dagger_{-p} \alpha_{-p} - 1 \right) \frac{\epsilon_p + p}{2\epsilon_p} - \left( \beta^\dagger_p \beta_p + \beta^\dagger_{-p} \beta_{-p} - 1 \right) \frac{\epsilon_p - p}{2\epsilon_p} \right] | \sigma \rangle \nonumber \\
&=& \lim_{\Lambda\to\infty}\int_{-\Lambda-q/2}^{\Lambda-q/2} \frac{dp}{2\pi} \left[ n_F(E_\alpha) - n_F(E_\beta) \right] + \frac{q}{2\pi}.
\eeqa
The first term represents the normal magnetization, which counts the difference of the number of the up- and down-spin particles 
and the second term denotes the anomalous magnetization. The anomalous magnetization corresponds to the coefficient of $h$ in $\mathcal{L}_{\rm ano}$ by putting $\varphi = qx$ in Eq.~(\ref{ano}).
On the other hand, the LO state does not have the anomalous magnetization
because there is no phase degree of freedom\footnote{It does not necessarily imply the absence of magnetization in the LO phase \cite{machida}.}.




 
 


      
      

\subsection{2fNJL$_2$ case}

However, it is not evident whether the same features hold for other models such as the 2fNJL$_2$ model (\ref{2fnjl}), which is an anomaly-free model. We shall see the different features for the 2fNJL$_2$ model and how anomaly is responsible to these differences.
For this model, the duality transformation may be modified as $\psi \rightarrow \chi = \frac{1}{2}(1-\gamma^5\tau_3)\psi + \frac{1}{2}(1+\gamma^5\tau_3)\psi^*$ to include the flavor dependence\footnote{One may consider the flavor independent transformation,$\psi \rightarrow \chi = \frac{1}{2}(1-\gamma^5)\psi + \frac{1}{2}(1+\gamma^5)\psi^*$, but the resultant Lagrangian explicitly violates particle number conservation.
Therefore we treat only the flavor dependent transformation in the following.},
so that the Lagrangian (\ref{2fnjl}) can be written as,
\beqa
 {\cal L} = {\bar \chi}i\ds\partial \chi + \frac{G}{2} \left[ \left(\bar{\chi}^c\tau_3 \chi \right) \left( \bar{\chi}^c\tau_3 \chi \right)^\dagger  + \left({\bar\chi}i\gamma^5\tau_1\chi\right)^2 + \left({\bar\chi}i\gamma^5\tau_2\chi\right)^2 \right].
\eeqa
For the chemical potential term, it is changed to,
\beq
\delta L= -h\left(\chi_L^{u*}\chi_L^u - \chi_R^{u*}\chi_R^u - \chi_L^{d*}\chi_L^d + \chi_R^{d*}\chi_R^d\right).
\eeq
Accordingly we have the Hamiltonian within the mean-field approximation by assuming $\left\langle{\bar\chi}i\gamma^5\tau_{1,2}\chi\right\rangle \equiv 0$ and
\beqa
  \Delta = - \frac{G}{2} \langle \bar{\chi}^{c}\tau_3 \chi \rangle (\neq 0),
\eeqa
for the charge-neutral system,
\beqa
 H_{2f} 
 &=& \int dx \left[ \tilde{\Psi}^\dagger \left( \hat{p} \sigma_3 + h + \sigma_1{\rm Re}\Delta + \sigma_2{\rm Im}\Delta \right)\tilde{\Psi} + \tilde{\Phi}^\dagger \left( \hat{p} \sigma_3 + h + \sigma_1{\rm Re}\Delta - \sigma_2{\rm Im}\Delta \right)\tilde{\Phi} + \frac{|\Delta|^2}{2G} \right], \label{ham2f}
\eeqa
where, $\tilde{\Psi}^\dagger = (\chi^u_R, \chi^{u*}_L)$, $\tilde{\Phi}^\dagger = (\chi^{d*}_R, \chi^d_L)$.

If there is only the $u$-quark sector, the Hamiltonian is reduced to the one-flavor case (\ref{ham1f}).
In the LO state or in the case of no phase factor in the gap function, two Hamiltonians of the $\tilde{\Phi}$ and $\tilde{\Psi}$ sectors become identical
and the total Hamiltonian is reduced to the one-flavor case except the overall factor.
Therefore the phase diagram of the LO state is not changed for any number of flavor.

\begin{figure*}[]
\vspace{1cm}
 \centering
 \begin{tabular}{c}
      \begin{minipage}{0.5\hsize}
        \begin{center}
          \includegraphics[width=8cm]{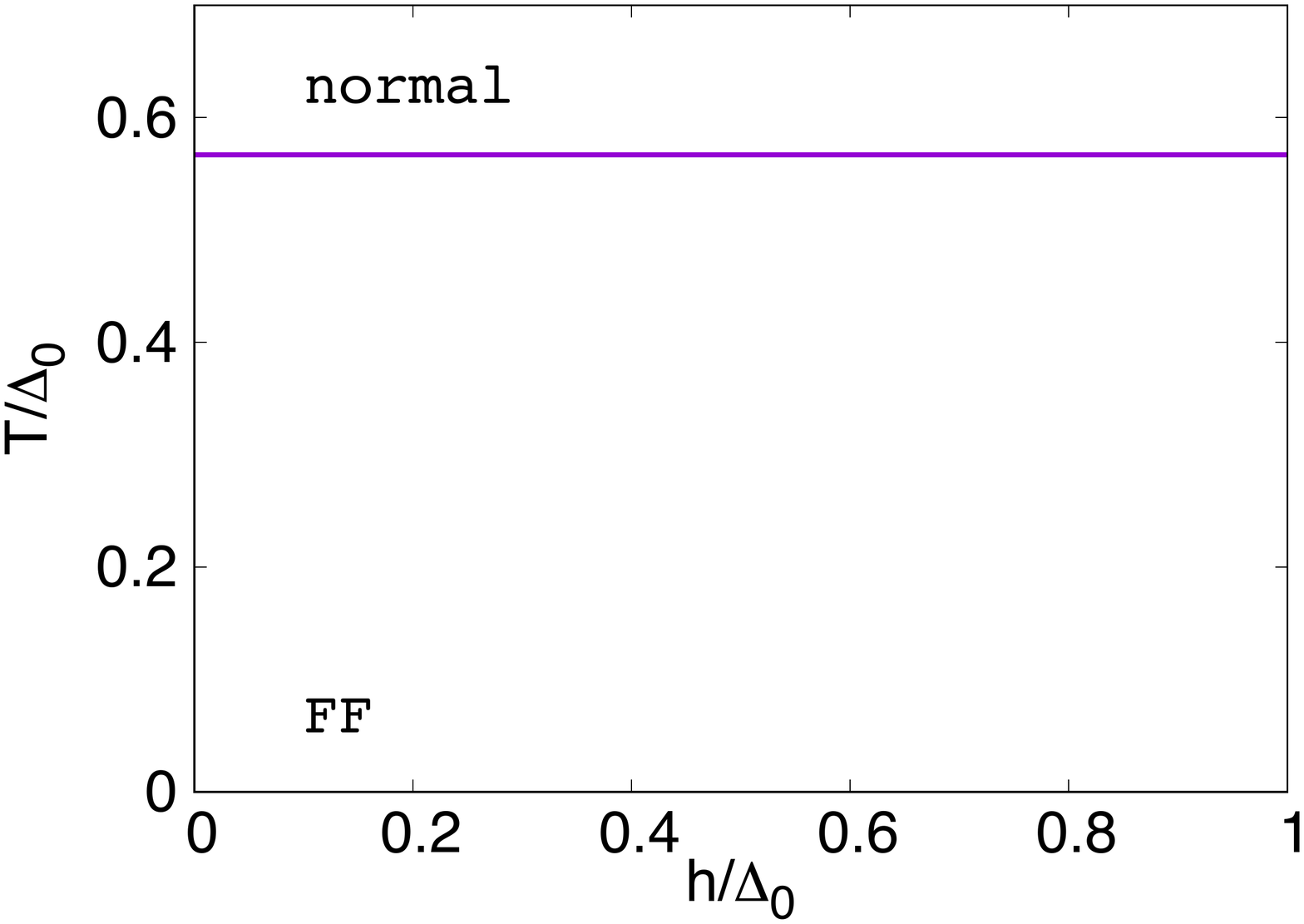}
        \end{center}
      \end{minipage}

      \begin{minipage}{0.5\hsize}
        \begin{center}
          \includegraphics[width=8cm]{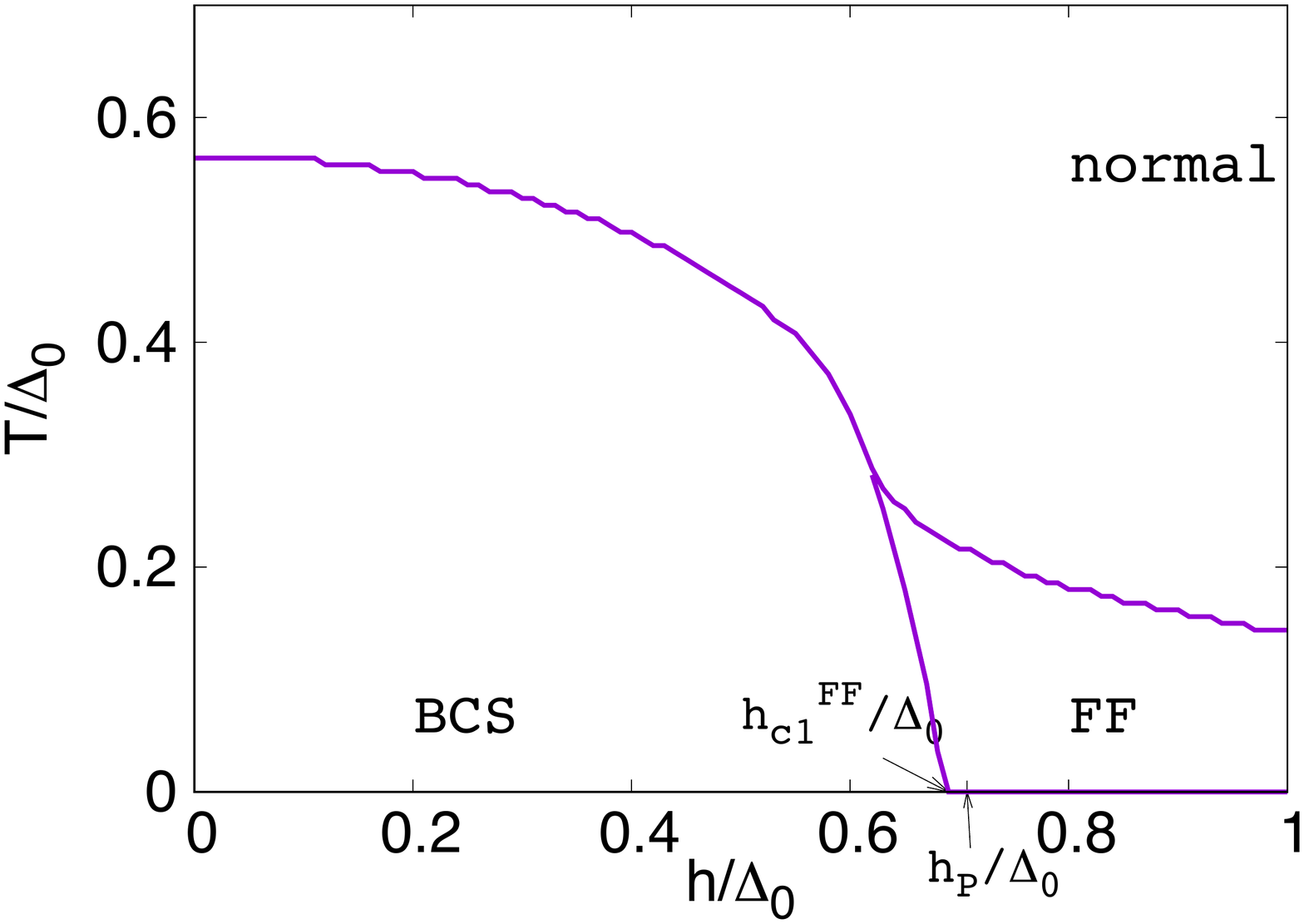}
        \end{center}
      \end{minipage}      
 \end{tabular}
 \caption{The difference of the regions of the FF phase between the one-flavor (left panel) and two-flavor (right panel) cases.
 $\Delta_0$ denotes the magnitude of the gap function $\Delta$ at $T=h=0$.
 $h_P$ represents the Pauli paramagnetic limit, $h_P/\Delta_0 = 1/\sqrt{2}$,
 and $h_{c1}^{FF}$ denotes the lower critical field at $T=0$, where the first order phase transition occurs, $h_{c1}^{FF}/\Delta_0 \simeq0.68$ in the 2fNJL$_2$ model, while $h_{c1}^{FF}=0$ in the NJL$_2$ model.
The higher critical field $h_{c2}^{FF}$ diverges at $T=0$ in both cases due to the perfect nesting (see the text).}
 \label{fig1}
 \vspace{0.5cm}
\end{figure*}

However, for the FF state, the phase diagram is different between one- and two-flavor cases due to the existence of anomaly.
Actually the Fig.\,\ref{fig1} shows the difference of the appearing region of the FF phase between the one- and two-flavor cases.
In the two-flavor case,  the $u$-quark sector has the energy spectrum (\ref{spect})
while the $d$-quark sector has the similar energy spectrum with the opposite sign of $q$.
Since the magnetization is given by summing up both contributions of the $u$- and $d$-sectors, they completely cancel each other.
This may be also infered from the anomaly relation in the flavor-symmetric matter,
\beq
\partial_\mu j^{\mu} =\frac{1}{4\pi}{\rm tr}(\tau_3)\epsilon_{\mu\nu}C^{\mu\nu},
\eeq
for {$j^\mu={\bar\chi}\gamma^\mu\chi$} in the 2fNJL$_2$ model.
Consequently the phase diagram for RKC is the same as the one given by Machida and Nakanishi \cite{machida} for the LO state, once chemical potential is replaced by the magnetic field;
they studied the possibility of the FFLO state in the quasi-one dimensional system by changing the strength of the magnetic field.
They used the linear dispersion approximation near the Fermi surface, so that there appear Dirac electrons with definite motions, the light and right moving electrons for each spin state.
Solving the Bogoliubov-de Gennes equation self-consistently within the mean-field approximation, they found the FFLO state above the critical magnetic field.
They also found that the phase boundaries between the FF and LO states and the normal phase are identical as they should be.
For the 2fNJL$_2$ model, the Hamiltonian (\ref{ham2f}) looks identical with the one argued by Machida and Nakanishi with the following correspondence,
\beqa
  \chi^u_R &\leftrightarrow& \psi_\downarrow, \nonumber \\
  \chi^u_L &\leftrightarrow& \phi_\uparrow, \nonumber \\
  \chi^d_R &\leftrightarrow& \psi_\uparrow, \nonumber \\
  \chi^d_L &\leftrightarrow& \phi_\downarrow,
\eeqa
where $\psi(\phi)$ represents the left(right) moving electron field and the up(down) arrow denotes the up(down) spin state.


\section{Nesting for iCP}

Nesting of the Fermi surface is one of the important concepts in condensed matter physics \cite{cdw,sdw,shimahara}. 
It is based on the geometry of the Fermi surface and almost free from dynamics: 
typical example is charge density wave or spin density wave in quasi-one dimensional systems.
As is already mentioned, the nesting effect is most prominent at $T=0$. So we, in the following, concentrate on the low temperature case. 
Nesting may be also a driving mechanism for iCP.
It has been sometimes discussed that chiral spiral appears due to nesting in 1+1 dimensions,
because there is opened an energy gap $m$ at the Fermi surface of massless quarks
and the wave number $q$ takes $2\mu$ at the same time.
On the other hand it looks rather difficult to interpret the onset of RKC by nesting,
because the wave number takes zero at the threshold.
We'd like to give some remarks about the relation between iCP and nesting.    

First of all we point out that it is too naive for the onset of chiral spiral to be ascribed to nesting.
We have seen that axial anomaly plays an important role for the relation $q=2\mu$.
Moreover, chiral spiral develops for arbitrary chemical potential below the critical temperature.
These are peculiar consequences within the NJL$_2$ model.
Actually we have seen in the 2fNJL$_2$ model that there exists a critical chemical potential $\mu_c$, above which chiral spiral develops.
The phase transition is of the first order in this case,
and the wave number takes a finite value of ${\mathcal O}(2\mu)$ at $\mu_c$.
Interestingly, the wave number takes the same order of magnitude as in the NJL$_2$ model.
Note that the magical relation $q=2k_F$ for nesting in 1+1 dimensions has been derived by the lowest-order perturbation;
e.g., the Lindhard function, which is the lowest order density-density correlation function or susceptibility, logarithmically diverges at $q=2k_F$ at $T=0$ in 1+1 dimensions to lead to formation of density wave \cite{grud}.
In the present model, $k_F$ means the Fermi momentum of the no-interacting quarks, that is, $k_F=\mu$.
In our case the energy gap is generated by the non-perturbative effect and the magical relation may not hold as it does.
On the other hand, we can see that the number of the wave number approaches $2k_F$ at $T=0$ around the critical chemical potential for the transition to the chiral-restored phase, where the non-perturbative effect becomes tiny and the perturbative result should hold.
Thus we can see that nesting may play an important role for chiral spiral.

For RKC the phase transition is of the second order and the wave number takes $q=0$ at the critical chemical potential $\mu_c^{\rm RKC}=2/\pi$ \cite{thies03,nic}.
However, the number of the wave number rapidly increases in the RKC phase and immediately approaches $q={\mathcal O}(2\mu)$.
Thus one may say nesting works except a small region around $\mu_c^{\rm RKC}$.




It should be interesting to see how such nesting effect manifests after the duality transformation.
Since the Hamiltonian describes a kind of superconducting phase, a different kind of nesting should be seen. 
There are two kinds of nesting:\,one (type-I) is familiar as a driving mechanism of charge density wave or spin density wave in quasi-one dimensional system \cite{sdw,cdw,grud}.
The other one is responsible to the FFLO state (type-I\hspace{-.1em}I).
In the magnetic field, two Fermi spheres with different Fermi momenta $p_F^i$ are created by the paramagnetic effect,
if any interaction is absent.
Nesting in the type-I\hspace{-.1em}I case is a combination of the inversion and translation of one Fermi sphere by $\delta p_F\equiv |p_F^1-p_F^2|$ to match with another one. 
In particular we shall see that RKC can be more easily understood by the type-I\hspace{-.1em}I nesting. 

Note that the FFLO state is not necessarily induced in the presence of the magnetic field.
Instead there is a competition between the paramagnetic effect and the Cooper pairing effect;
the paramagnetic effect favors a specific spin state and leads to the difference of the Fermi momenta of the two spin states,
while the Cooper pairing effect becomes maximum for the equal Fermi momenta \cite{sar}.
When the paramagnetic effect dominates over the Cooper pairing effect,
the FFLO state is realized due to the type-I\hspace{-.1em}I nesting. 
The landmark of the lower critical filed is then given by the Pauli paramagnetic (Chandrasekhar-Clogston) limit, $h_P/\Delta_0=1/\sqrt{2}$ \cite{sar}.

\begin{figure*}[t]
 \centering
 \begin{tabular}{c}
 
      \hspace{-2.0cm}
 
      \begin{minipage}{0.5\hsize}
        \begin{center}
          \includegraphics[width=6cm]{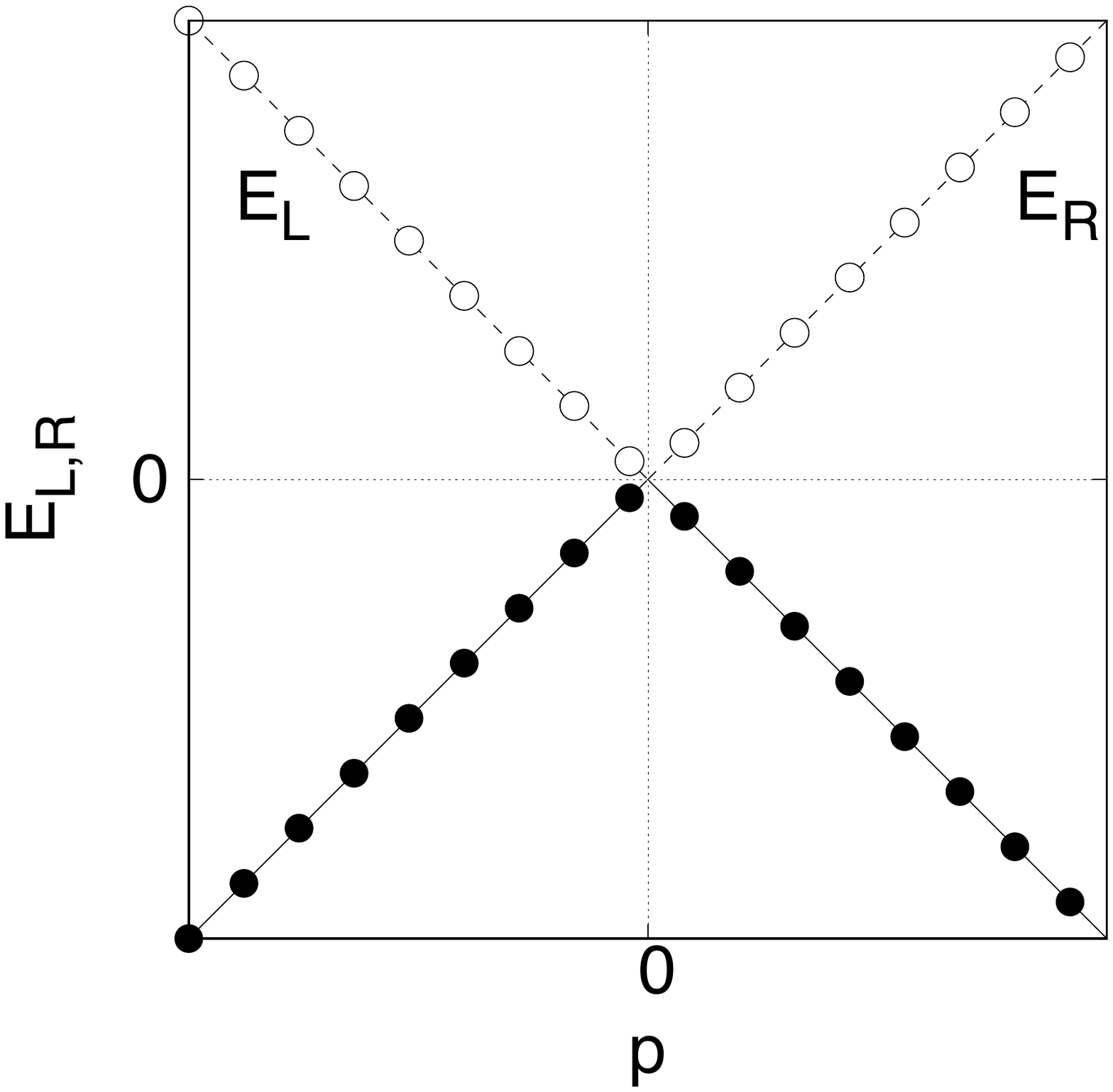}
          \vspace{0.7cm}
        \end{center}
      \end{minipage}

   \hspace{-3.0cm}
 
       \begin{minipage}{0.5\hsize}
        \begin{center}
          \includegraphics[width=6cm]{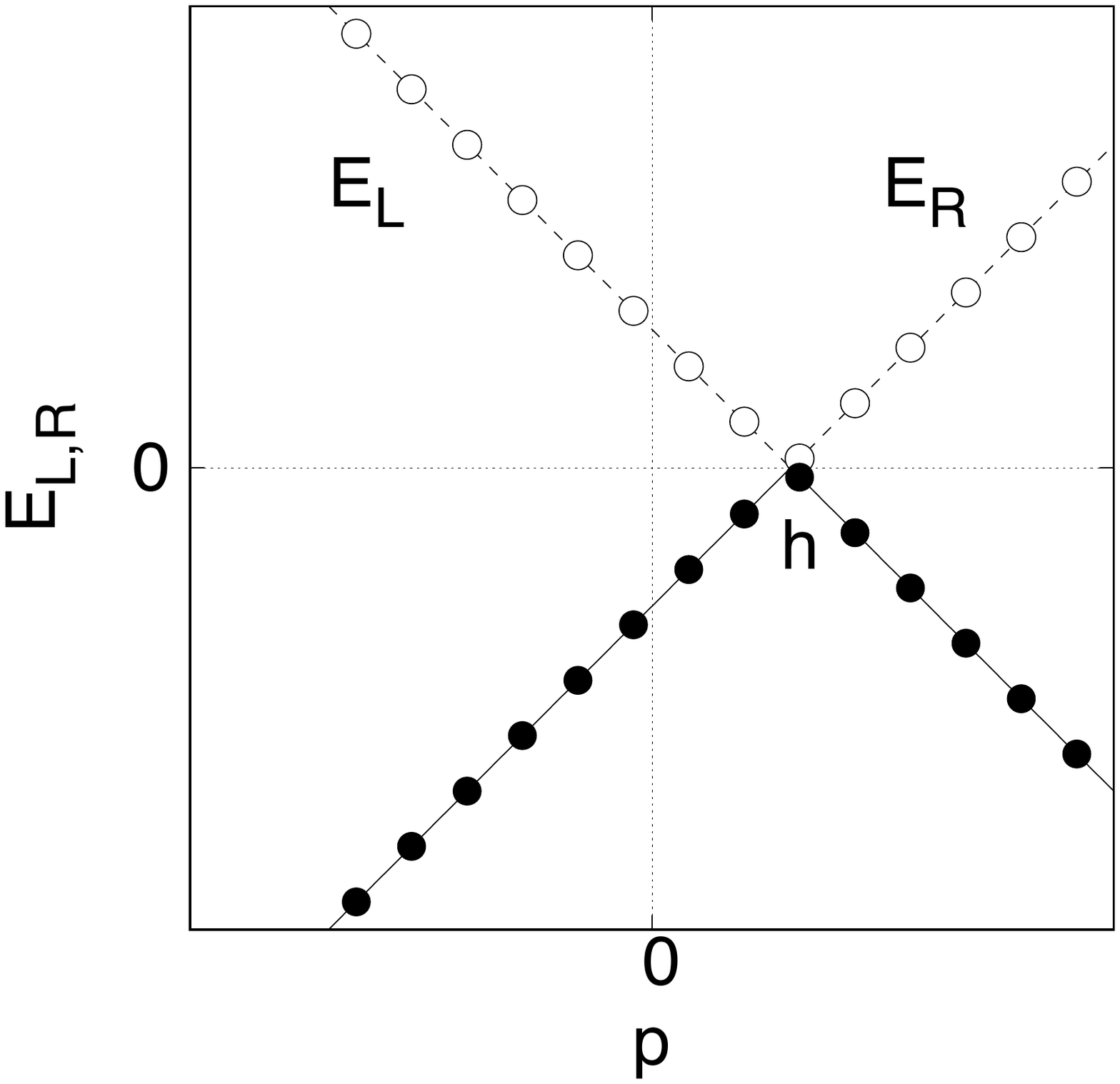}
          \hspace{10cm}
          \vspace{0.7cm}
        \end{center}
      \end{minipage} 
      
     \hspace{-3.0cm}
     
     \begin{minipage}{0.5\hsize}
        \begin{center}
          \includegraphics[width=6cm]{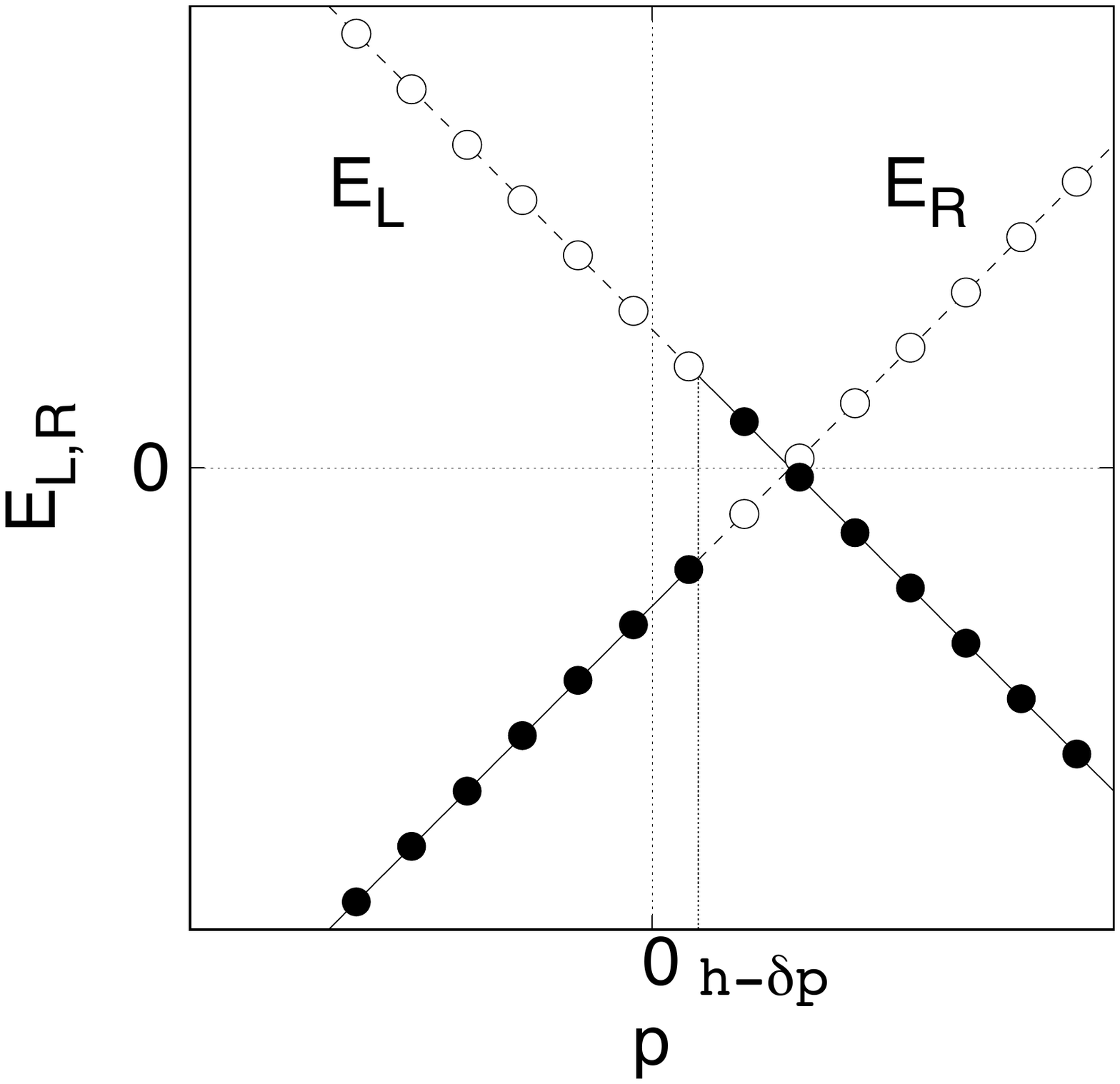}
          \vspace{0.7cm}
        \end{center}
      \end{minipage}
      
 \end{tabular}
 \caption{The vacua and an excited state with or without the magnetic field in the absence of the Cooper pairing.
Filled(unfilled) circles denote the occupied(unoccupied) states.
The left panel shows the normal vacuum in the absence of the magnetic field with the energy spectrum, $E_{R,L} = \pm p$.
The midlle panel shows the $h$-vacuum to give the energy spectrum, $E_R = p - h, E_L = -p + h$.
The right panel shows an excited state where the number density of L-particles is generated by $\frac{\delta p}{2\pi}$ and that of right-handed holes is generated by $\frac{\delta p}{2\pi}$ compared to the $h$-vacuum.}
 \label{q5}
 \vspace{0.5cm}
\end{figure*}

We can see by two steps how the type-I\hspace{-.1em}I nesting works by considering the energy spectra given in Fig.\,\ref{q5}.
Normal vacuum is constructed by filling the negative energy states as given by the left panel in Fig.\,\ref{q5}.
In the first step we consider the paramagnetic effect.
When the magnetic field is applied to the normal vacuum,
the energy spectra are changed for L- and R-particles.
The middle panel in Fig.\,\ref{q5} shows the vacuum in the presence of the magnetic field ($h$-vacuum), where all the negative energy states are occupied to make the total energy to be minimum.
In the $h$-vacuum, there is an imbalance between the number of R- and L-particles due to the paramagnetic effect.
Therefore magnetization can be evaluated to be $M_h = h / \pi$ because the number density of R-particles is increased by $\frac{h}{2\pi}$ and that of L-particles is inversely reduced by $\frac{h}{2\pi}$ compared to the normal vacuum.
We can also consider the excited states where some particle-holes are generated from the $h$-vacuum.
The right panel in Fig.\,\ref{q5} shows an excited state where the number density of particle-hole pairs is evaluated as $\frac{\delta p}{2\pi}$ by using the energy-level spacing $2\pi/L$.
In the excited state, the number density of R-particles is reduced by $\frac{\delta p}{2\pi}$ and that of L-particles is increased by $\frac{\delta p}{2\pi}$ compared to the $h$-vacuum as shown in the middle panel in Fig.\,\ref{q5}. 
Consequently, magnetization takes the finite value, $M = (h-\delta p)/\pi$.
Note that we cannot choose the optimal one among them in this step,
because the basic variational principle should be applied to the total energy after taking into account the Cooper pairing effect.

\begin{figure*}[]
 \centering
 \begin{tabular}{c}

      \begin{minipage}{0.5\hsize}
        \begin{center}
          \includegraphics[width=8cm]{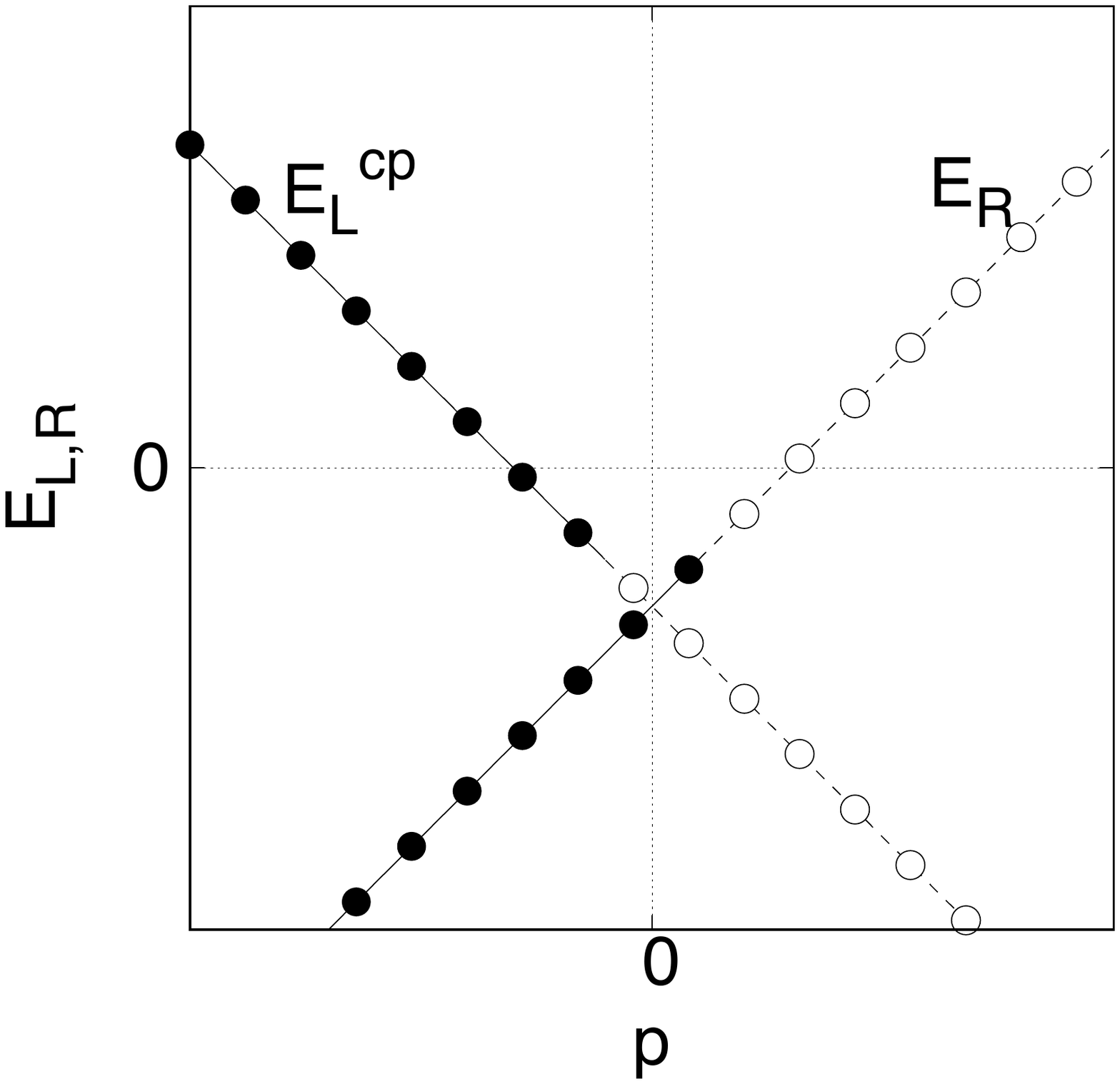}
          \vspace{0.7cm} 
        \end{center}
      \end{minipage}
      
      \hspace{-2.0cm}
      
      \begin{minipage}{0.5\hsize}
        \begin{center}
          \includegraphics[width=8cm]{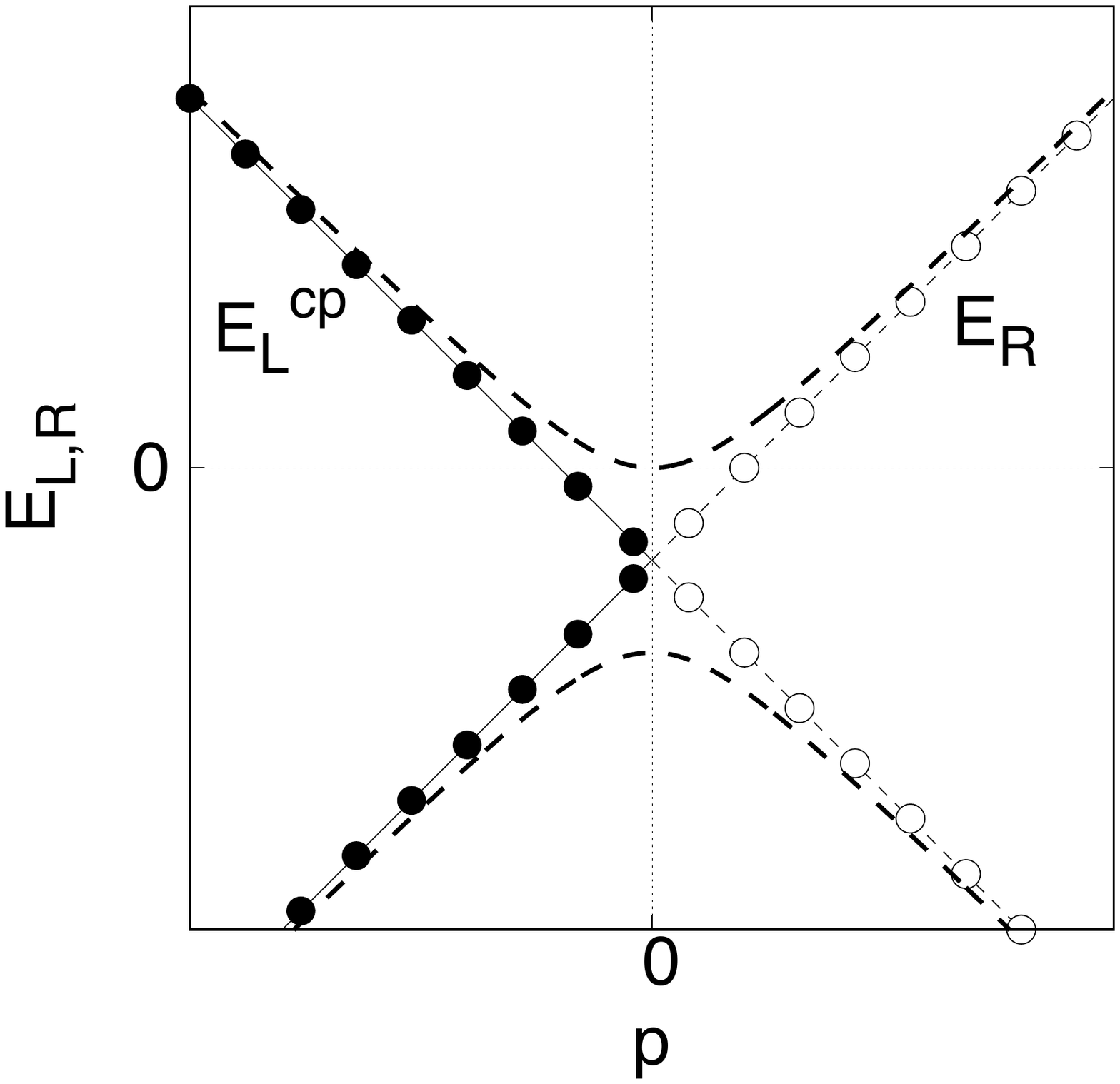}
          \vspace{0.7cm}
        \end{center}
      \end{minipage}      
      
 \end{tabular}
 \caption{Configuration of the Fermi surface at the transition point from the normal phase to the superconducting phase.
Filled(unfilled) circles denote the occupied(unoccupied) states. 
The left panel shows inversion spectrum of $E_L$, $E^{cp}_{L}$, from the right panel in Fig.\,\ref{q5}.
The right panel shows the way of the pairing after the momentum shift $\pm (h - \delta p)$ to give the quasiparticle energy spectra ($E_\alpha$ and $E_\beta^c$) denoted by the bold dashed lines.}
 \label{super}
 \vspace{0.5cm}
\end{figure*}

In the second step we consider the Cooper pairing effect.
In Fig.\,\ref{super} the construction of the quasi-particle energy is graphically explained:\,by the inversion of one energy spectrum and the relative momentum shift with $q=2(h - \delta p)$ for the excited state, we arrive at the quasiparticle energy with the pairing gap at the Fermi surface.
Such momentum shift corresponds to the wave number of the spatial modulation of the gap function.
When $q \neq 0$ takes the energy minimum,
the FF state appears in the ground state in place of the BCS state.
When $\delta p = h$, we can see the usual BCS gap at the Fermi surface $p_F=0$ by inversion of one spectrum,
so that the gap function is constant.
On the other hand, when the same manipulation is applied for the $h$-vacuum,
we can see the momentum must be shifted by $q=2h$ after inversion.

The $h$-vacuum is realized at the phase boundary between the FFLO state and the normal phase.
The typical momentum $q=2h$ can be also seen by considering the correlation function between the Cooper pairs in the normal phase:
it depends on the dimensionality and logarithmically diverges at $q=2h$ in 1+1 dimensions \cite{tho,nsr,oha}.

We consider how the type-I nesting is mapped to the type-I\hspace{-.1em}I nesting by the duality transformation.
As is discussed above axial anomaly sometimes conceals the nesting effect,
we first discuss it by using an anomaly-free model, such as the 2fNJL$_2$ model.
After applying the duality transformation, we consider the FFLO state under the magnetic field in the vacuum, described by the Hamiltonian (\ref{ham2f}).
Our model then becomes the same one discussed by Machida and Nakanishi in the context of condensed matter physics.
Accordingly the phase diagram becomes the same.
For the LO state, the phase transition is of the second order from the BCS state at the lower critical field $h_{c1}^{\rm LO}$, $h_{c1}^{\rm LO}/\Delta_0=2/\pi$. 
The wave number increases from the zero value, which reflect the type-I\hspace{-.1em}I nesting.
The excited state, $\delta p = h$, is realized at the phase boundary between the BCS state and the LO state.
On the other hand, for the FF state, the phase transition is of the first order with finite wave number of ${\mathcal O}(2h)$.
This feature looks somewhat different from the LO state,
but one may see the type-I\hspace{-.1em}I nesting works except the small region of the lower critical field $h_{c1}^{\rm FF}$, $h_{c1}^{\rm FF}/\Delta_0\simeq 0.68$.
We can also see that the phase boundaries from the LO and FF phases to normal phase coincide with each other \cite{yoshiike3}.
Thus we can say the type-I\hspace{-.1em}I nesting works for the FFLO state.
It should be interesting to note that the upper critical field $h_{c2}$ diverges for both phases as $T\rightarrow 0$ in 1+1 dimensions due to the perfect type-II nesting with $q=2h$.





For the NJL$_2$ model, the argument about the LO state is unchanged, since the order parameter is real and the model Hamiltonian is reduced to (\ref{ham1f}), while that about the FF state is greatly modified. 
We shall discuss some surprising aspects of the FF state in the NJL$_2$ model.
The FF state appears once a tiny magnetic field is applied due to anomaly inducing the anomalous term ${\mathcal L}_{\rm ano}$; thermodynamic potential includes the linear term of $q$,
\beqa
 \Omega 
 &=& -\frac{\Lambda^2}{2\pi} - \frac{m^2}{2\pi} \ln(2\Lambda/m) - \frac{m^2}{4\pi} + \frac{1}{2\pi}(h-q/2)^2 - \frac{h^2}{2\pi}+ \frac{m^2}{2G}.
\eeqa
The minimum condition gives rise to the relation, ${q = 2h}$, even if the magnetic field is tiny (see Appendix B for detail).
Hence the lower critical field becomes zero and the perfect type-II nesting always holds with $q=2h$.











\section{Summary and concluding remarks}

Using the duality relation we have discussed the roles of axial anomaly and nesting in the context of iCP.
We have seen that the NJL$_2$ model has $U(1)_L\times U(1)_R$ symmetry in the classical level,
but symmetry is broken due to axial anomaly in the presence of the gauge field, $U(1)_L\times U(1)_R\rightarrow U_V(1)$.
Invoking the technique of the fictious gauge field, $B^\mu=(\mu,0)$, 
such anomaly effect can be built in the thermodynamic potential in medium as anomalous quark number, which is given by spectral asymmetry of the quark field \cite{tatsumi}.
The NJL$_2$ model can be written as another form by way of the duality transformation.
New Lagrangian has a suitable form to describe 
a kind of superconductivity in the presence of the magnetic field, which resembles the FFLO state in the condensed matter physics. 

After the duality transformation we have seen a different manifestation of axial anomaly:
spectral asymmetry of quasiparticles does not necessarily implies the anomalous number in this case, as has been explicitly shown.
Instead, anomalous magnetization is generated for the complex order parameter.
Existence of magnetization means the different numbers for L- and R-quarks and leads to the different sizes of the Dirac seas.
Consequently, the FF state distinctively behaves in the magnetic field due to anomaly,
and the phase diagram becomes much different from the one for the FF state in condensed matter physics.
It develops, once the magnetic field is applied, i.e., the lower critical field $h_{c1}^{\rm FF}=0$.
We have confirmed this result by considering an anomaly free-model, two flavor NJL$_2$ model, 
where the FF state appears beyond the lower critical field $h_{c1}$ as in condensed matter physics.

Based on these considerations we have discussed how nesting plays in the context of iCP.
In the case of the anomaly free model we have first seen that the usual nesting (type-I nesting) works for both chiral spiral and RKC; the wave number becomes ${\mathcal O}(2\mu)$.  After the duality transformation we have considered the different type of nesting (type-I\hspace{-.1em}I nesting) in the context of superconductivity.
Using the concept of the type-I\hspace{-.1em}I nesting, we have shown that the type-I\hspace{-.1em}I nesting holds for both cases.
Interestingly, we have observed an ideal type-I\hspace{-.1em}I nesting for the LO state, where the new phase is brought about by the second order phase transition,
and the wave number increases from the zero value to the maximum value of $2\mu$.
For the NJL$_2$ model, anomaly modifies these pictures, especially for chiral spiral.
Sometimes one may attribute the relation $q=2\mu$ to the type-I nesting,
but we have emphasized that axial anomaly may be mainly responsible to this relation to conceal the nesting effect: nesting effect should be really appreciated in anomaly-free models.  

Finally we'd like to make a comment about a phenomenological perspective of our result. 
It is a possibility of a new type of superconductivity in low dimensional systems in condensed matter physics, 
which corresponds to our FF state and reflects anomaly.
If it can be created, we shall see the FF state for a tiny magnetic field.

We have treated Lagrangians in the chiral limit here, but the effect of current mass should be included for a realistic discussions \cite{kar,yos2}. The extention to flavor $SU(3)$ should be also interesting \cite{mor}. These subjects are left for future studies.

\begin{acknowledgments}
 The authors thank T.-G. Lee and N. Yasutake for their interest in this work and useful discussions. We also thank members of Nuclear Theory group at Kyoto University for discussions.
 This work is partially supported by Grants-in-Aid for Japan Society
 for the Promotion of Science (JSPS) fellows No.27-1814.
\end{acknowledgments}

\appendix
\section{Consistent UV regularization}
We explain the UV cutoff procedure in the calculation of the physical quantities such as quark number density or magnetization.
In the free theory without the magnetic field $h$, the fermion field renders by the chiral description,
\beqa
 \chi^{(0)}(x) =  \int \frac{dp}{2\pi}e^{ipx} 
 \begin{pmatrix}
  a^{(0)}_p \theta(p) + b^{(0)\dagger}_{-p} \theta(-p) \\
  a^{(0)}_p \theta(-p) - b^{(0)\dagger}_{-p} \theta(p)
 \end{pmatrix},
\eeqa
where  $a^{(0)}_p(b^{(0)}_p)$ denotes the annihilation operator of the (anti-)particle with energy spectrum $E=|p|$.
When $h$ is switched on, we define the creation and annihilation operators,
\beqa
  \chi(x) =  \int \frac{dp}{2\pi}e^{ipx} 
 \begin{pmatrix}
  a_p \theta(p) + b^\dagger_{-p} \theta(-p) \\
  a_p \theta(-p) - b^\dagger_{-p} \theta(p)
 \end{pmatrix}, \label{chitilde}
\eeqa
where $a_p(b_p)$ denotes the annihilation operator of the (anti-)particle with $E=(p-h){\rm sign}(p)$.
It seems that the energy spectrum of R(L)-particles and anti-particles is just shifted by $-h(h)$ from no magnetic field case.
The $h$-vacuum $|0\rangle$ is defined by filling the all ``negative energy states",
\beqa
 a_p |0\rangle &=& 0 ~~(p>h,\,p<0), \\
 a^\dagger_p |0\rangle &=& 0 ~~(0<p<h), \\
 b_{-p} |0\rangle &=& 0 ~~(p>h,\,p<0), \\
 b^\dagger_{-p} |0\rangle &=& 0 ~~(0<p<h).
\eeqa
In this case, the quark number density can be calculated as,
\beqa
 n_0 &\equiv& \frac{1}{2} \int \frac{dx}{L} \langle 0 | [\chi^\dagger, \chi] | 0 \rangle \nonumber \\
 &=& \lim_{\Lambda\to\infty} \langle 0 | \bigg[\int_0^\Lambda \frac{dp}{2\pi} \left( a^\dagger_p a_p - \frac{1}{2} \right) - \int^0_{-\Lambda} \frac{dp}{2\pi} \left( b^\dagger_{-p} b_{-p} - \frac{1}{2} \right) \nonumber \\
 && + \int^0_{-\Lambda} \frac{dp}{2\pi} \left( a^\dagger_p a_p - \frac{1}{2} \right) - \int_0^\Lambda \frac{dp}{2\pi} \left( b^\dagger_{-p} b_{-p} - \frac{1}{2} \right) \bigg] | 0 \rangle \nonumber \\
 &=& \lim_{\Lambda\to\infty} \left[\int^h_0 \frac{dp}{2\pi} \langle 0 | a^\dagger_p a_p | 0 \rangle - \int^h_0 \frac{dp}{2\pi} \langle 0 | b^\dagger_{-p} b_{-p} | 0 \rangle \right]  \nonumber \\
 &=& 0, \label{magtil}
\eeqa
which means that the number density of R-particles is $\frac{h}{2\pi}$
and that of L-anti-particle is simultaneously produced by just $\frac{h}{2\pi}$ compared to the normal vacuum.
Therefore the net quark number density vanishes
but magnetization has the finite value, $\frac{h}{\pi}$.
It corresponds to the right panel in Fig.\,\ref{q5}.
Furthermore the momentum cutoff $\Lambda$ is introduced to regularize the divergence.

Next we consider the quark number density in the FF state.
By equating Eq.\,(\ref{chitilde}) with Eq.\,(\ref{chi}), the Bogoliubov transformation is obtained as,
\beqa
 \alpha_p &=& \frac{1}{\sqrt{2\epsilon_p}}\left\{ \begin{array}{lc} 
                         a_{p+q/2}\sqrt{\epsilon_p + p} - a^\dagger_{-p+q/2} \sqrt{\epsilon_p - p} & (p>q/2) \\
                         a_{p+q/2} \sqrt{\epsilon_p + p} + b_{p-q/2} \sqrt{\epsilon_p - p} & (-q/2<p<q/2) \\
                         b^\dagger_{-p-q/2} \sqrt{\epsilon_p + p} + b_{p-q/2} \sqrt{\epsilon_p - p} & (p<-q/2)
                         \end{array} \right. ,\\
 \beta_p &=& \frac{1}{\sqrt{2\epsilon_p}}\left\{ \begin{array}{lc} 
                         b_{p-q/2} \sqrt{\epsilon_p + p} - b^\dagger_{-p-q/2} \sqrt{\epsilon_p - p} & (p>q/2) \\
                         a^\dagger_{-p+q/2} \sqrt{\epsilon_p + p} - b^\dagger_{-p-q/2} \sqrt{\epsilon_p - p} & (-q/2<p<q/2) \\
                         a^\dagger_{-p+q/2} \sqrt{\epsilon_p + p} + a_{p+q/2} \sqrt{\epsilon_p - p} & (p<-q/2)
                         \end{array} \right. .
\eeqa
Setting $m=0$ in Eq.\,(\ref{chi}), it should reproduce the free field theory with $h$ even if $q$ is still finite.
Therefore the quark number density in the FF state becomes independent on $q$ and coincides with Eq.\,(\ref{magtil}) at $m=0$.
At $m=0$, the transformation renders,
\beqa
 \alpha_p &=& \left\{ \begin{array}{lc} 
                         a_{p+q/2} & (p>0) \\
                         b_{p-q/2} & (p<0)
                         \end{array} \right. , \label{bog1} \\
 \beta_p &=& \left\{ \begin{array}{lc} 
                         b_{p-q/2} & (p>q/2) \\
                         a^\dagger_{-p+q/2} & (0<p<q/2) \\
                         -b^\dagger_{-p-q/2} & (-q/2<p<0) \\
                         a_{p+q/2} & (p<-q/2)
                         \end{array} \right. . \label{bog2}
\eeqa
Once we introduce the lower and upper momentum cutoffs independently, $[\Lambda_{\rm min},\Lambda_{\rm max}]$, to determine the appropriate one in the FF state,
the quark number density renders from Eq.\,(\ref{chi}),
\beqa
 n &\equiv& \frac{1}{2} \int \frac{dx}{L} \langle \sigma | [\chi^\dagger, \chi] | \sigma \rangle \nonumber \\
 &=& \lim_{ \begin{array}{c} 
                    \Lambda_{\rm max} \to \infty \\
                    \Lambda_{\rm min} \to -\infty 
                   \end{array} }
                 \langle \sigma | \int_{\Lambda_{\rm min}}^{\Lambda_{\rm max}} \frac{dp}{2\pi} \left[ \left( \alpha^\dagger_p \alpha_p - \alpha^\dagger_{-p} \alpha_{-p} \right) \frac{\epsilon_p + p}{2\epsilon_p} + \left( \beta^\dagger_p \beta_p - \beta^\dagger_{-p} \beta_{-p} \right) \frac{\epsilon_p - p}{2\epsilon_p} \right] | \sigma \rangle.
\eeqa
Setting $m=0$, the ground state becomes $|0\rangle$ and the quark number density can be calculated as,
\beqa
 n &=& \lim_{ \begin{array}{c} 
                    \Lambda_{\rm max} \to \infty \\
                    \Lambda_{\rm min} \to -\infty 
                   \end{array} }
                    \langle 0 | \int_{\Lambda_{\rm min}}^{\Lambda_{\rm max}} \frac{dp}{2\pi} \left[ \left(ß \alpha^\dagger_p \alpha_p - \alpha^\dagger_{-p} \alpha_{-p} \right) \theta(p) + \left( \beta^\dagger_p \beta_p - \beta^\dagger_{-p} \beta_{-p} \right) \theta(-p) \right] | 0 \rangle \nonumber \\
 &=& \lim_{ \begin{array}{c} 
                    \Lambda_{\rm max} \to \infty \\
                    \Lambda_{\rm min} \to -\infty 
                   \end{array} }
  \langle 0 | \left[\int_0^{\Lambda_{\rm max} + q/2} \frac{dp}{2\pi} \left( a^\dagger_p a_p - b^\dagger_{-p} b_{-p} \right) + \int^0_{\Lambda_{\rm min} + q/2} \frac{dp}{2\pi} \left( a^\dagger_{p} a_{p} - b^\dagger_{-p} b_{-p} \right) \right] | 0 \rangle,
\eeqa
where the transformation (\ref{bog1}), (\ref{bog2}) is used.
Therefore we can see that quark number density in the FF state reproduces Eq.\,(\ref{magtil}) at $m=0$
when the momentum cutoff is put to be asymmetric, $\Lambda_{\rm max}=\Lambda-q/2, \Lambda_{\rm min} = -\Lambda - q/2$.
It can be also confirmed that magnetization is consistently calculated in the same way by using that asymmetric cutoff.

\section{Thermodynamic potential in the FF state}
We calculate the thermodynamic potential in the FF state for the NJL$_2$ model.
To obtain the correct thermodynamic potential, the appropriate momentum cutoff,$[-\Lambda - q/2,\Lambda - q/2]$, should be used in the $\chi$ sector.
From the relation, $\chi^c = \gamma^5\chi^*$, the asymmetric cutoff in the $\chi^c$ sector should be $[-\Lambda + q/2,\Lambda + q/2]$.
From the Hamiltonian (\ref{ham-NG}), the thermodynamic potential at $T=0$ renders,
\beqa
  \Omega &\equiv& \frac{\langle \sigma | H_{1f} |\sigma\rangle}{L} \nonumber \\
 &=& \frac{1}{2} \int \frac{dx}{L} \langle \sigma |(\chi'^\dagger,\chi'^{c\dagger})
 \begin{pmatrix}
 -i\gamma^5\partial_x - \gamma^5(h - q/2) & \gamma^0 m \\
 \gamma^0 m & -i\gamma^5\partial_x + \gamma^5 (h - q/2)
 \end{pmatrix}
 \begin{pmatrix}
 \chi' \\
 \chi'^c
 \end{pmatrix}|\sigma\rangle + \frac{m^2}{2G} \nonumber \\
 &=& \frac{1}{2} \int^{\Lambda-q/2}_{-\Lambda-q/2} \frac{dp}{2\pi} \frac{1}{2\epsilon_p}\langle \sigma | \Big\{ [p-h+q/2] \left[(\epsilon_p + p) \alpha_p^\dagger \alpha_p + (\epsilon_p - p) \beta_p^c \beta_p^{c\dagger} \right] \nonumber \\
 &&~~~~~~~~~~~~~~~~~~~~~~~~~ - [p-h+q/2] \left[ (\epsilon_p + p) \alpha_p^c \alpha^{c\dagger}_p + (\epsilon_p - p) \beta_p^\dagger \beta_p \sin^2\theta \right] \Big\} |\sigma\rangle \nonumber \\
 &+& \frac{1}{2} \int^{\Lambda+q/2}_{-\Lambda+q/2} \frac{dp}{2\pi} \frac{1}{2\epsilon_p}\langle \sigma | \Big\{ [p+h-q/2] \left[ (\epsilon_p + p) \beta_p^\dagger \beta_p + (\epsilon_p - p) \alpha_p^c \alpha_p^{c\dagger} \right] \nonumber \\
 &&~~~~~~~~~~~~~~~~~~~~~~~~~ - [p+h-q/2] \left[ (\epsilon_p + p) \beta_p^c \beta^{c\dagger}_p + (\epsilon_p - p)\alpha_p^\dagger \alpha_p \right] \Big\} |\sigma\rangle \nonumber \\
 &+& m^2 \int^{\Lambda-q/2}_{-\Lambda+q/2} \frac{dp}{2\pi} \frac{1}{2\epsilon_p} \langle \sigma | \left( \alpha_p^\dagger \alpha_p - \beta_p^c \beta^{c\dagger}_p + \beta_p^\dagger \beta_p - \alpha_p^c \alpha_p^{c\dagger} \right) |\sigma\rangle + \frac{m^2}{2G}.
\eeqa
Assuming $|h-q/2|<m$, $\Omega$ can be reduced,
\beqa
 \Omega &=& -\int^{\Lambda-q/2}_{-\Lambda-q/2} \frac{dp}{2\pi} \frac{p}{\epsilon_p}(p-h+q/2) - m^2 \int^{\Lambda-q/2}_{-\Lambda+q/2} \frac{dp}{2\pi} \frac{1}{\epsilon_p} + \frac{m^2}{2G}. \nonumber \\
 &=& -\frac{\Lambda^2}{2\pi} - \frac{m^2}{2\pi} \ln(2\Lambda/m) - \frac{m^2}{4\pi} + \frac{1}{2\pi}(h-q/2)^2 - \frac{h^2}{2\pi}+ \frac{m^2}{2G} + {\mathcal O}(1/\Lambda).
\eeqa
The quadratic-divergence term is irrelevant and can be subtracted off.
The logarithmic divergence can be removed by the appropriate renormalization scheme independent of $q$ as in the GN model \cite{gro}.
We can see that $\Omega$ includes the linear term of $q$, so that  
there is always the minimum point at $q = 2h$.
In the case, $|h-q/2|>m$, the thermodynamic potential can be calculated in the same way.
However its energy minimum is larger than that at $q=2h$.

\end{document}